\documentclass[amsmath,aps,showpacs,a4paper,10pt]{revtex4}
 \usepackage{epsf}
 \usepackage{graphicx}    % Include figure files

\begin{document}

 \newcommand{\be}[1]{\begin{equation}\label{#1}}
 \newcommand{\ee}{\end{equation}}
 \newcommand{\bea}{\begin{eqnarray}}
 \newcommand{\eea}{\end{eqnarray}}
 \def\disp{\displaystyle}

 \def\gsim{ \lower .75ex \hbox{$\sim$} \llap{\raise .27ex \hbox{$>$}} }
 \def\lsim{ \lower .75ex \hbox{$\sim$} \llap{\raise .27ex \hbox{$<$}} }

 \begin{titlepage}

 \begin{flushright}
 arXiv:1312.1117
 \end{flushright}

 \title{\Large \bf Cosmological Applications of Pad\'e Approximant}

 \author{Hao~Wei\,}
 \email[\,email address:\ ]{haowei@bit.edu.cn}
 \affiliation{School of Physics, Beijing Institute
 of Technology, Beijing 100081, China}

 \author{Xiao-Peng~Yan\,}
 \affiliation{School of Physics, Beijing Institute
 of Technology, Beijing 100081, China}

 \author{Ya-Nan~Zhou}
 \affiliation{School of Physics, Beijing Institute
 of Technology, Beijing 100081, China}

 \begin{abstract}\vspace{1cm}
 \centerline{\bf ABSTRACT}\vspace{2mm}
 As is well known, in mathematics, any function could be
 approximated by the Pad\'e approximant. The Pad\'e approximant
 is the best approximation of a function by a rational function
 of given order. In fact, the Pad\'e approximant often gives
 better approximation of the function than truncating its
 Taylor series, and it may still work where the Taylor
 series does not converge. In the present work, we
 consider the Pad\'e approximant in two issues. First, we
 obtain the analytical approximation of the luminosity
 distance for the flat XCDM model, and find that the relative
 error is fairly small. Second, we propose several
 parameterizations for the equation-of-state parameter (EoS)
 of dark energy based on the Pad\'e approximant. They are well
 motivated from the mathematical and physical points of view.
 We confront these EoS parameterizations with the latest
 observational data, and find that they can work well. In
 these practices, we show that the Pad\'e approximant could
 be an useful tool in cosmology, and it deserves further
 investigation.
 \end{abstract}

 \pacs{98.80.-k, 95.36.+x, 98.80.Es, 98.80.Jk}

 \maketitle

 \end{titlepage}

 \renewcommand{\baselinestretch}{1.1}

%============================= section 1 ===================================

\section{Introduction}\label{sec1}

Since the great discovery of the current accelerated expansion
 of our universe, dark energy has become one of the most active
 fields in physics and astronomy~\cite{r1}. As is well known,
 Type Ia supernovae (SNIa) have been considered as a powerful
 probe to investigate this mysterious phenomenon~\cite{r1,r2}.
 We can constrain cosmological models by comparing their
 theoretical luminosity distances (or distance modulus
 equivalently) with the observational ones of SNIa. Therefore,
 computing the luminosity distance is fairly important in dark
 energy cosmology. On the other hand, it is also well known
 that the equation-of-state parameter (EoS) plays an important
 role in cosmology. The evolution of energy density of dark
 energy mainly depends on its EoS. Determining EoS of dark
 energy is one of the key tasks in cosmology~\cite{r1}.

As is well known, any function $f(x)$ can be approximated by
 the Taylor series expansion, namely, $f(x)=f(x_0)+f_1\,(x-x_0)
 +f_2\,(x-x_0)^2+\dots+f_n\,(x-x_0)^n$. Taylor expansion has
 been extensively used in many fields of physics. The so-called
 Pad\'e approximant can be regarded as a generalization of
 Taylor polynomial~\cite{r3,r4,r5,r6,r7}. In mathematics, a
 Pad\'e approximant is the best approximation of a function by
 a rational function of given order~\cite{r4}. In fact, the
 Pad\'e approximant often gives better approximation of the
 function than truncating its Taylor series, and it may still
 work where the Taylor series does not converge~\cite{r4}. For
 any function $f(x)$, its corresponding Pad\'e approximant
 of order $(m,\,n)$ is given by the rational
 function~\cite{r3,r4,r5,r6,r7}
 \be{eq1}
 f(x)=\frac{\alpha_0+\alpha_1 x+\cdots+\alpha_m x^m}{1+
 \beta_1 x+\cdots+\beta_n x^n}\,,
 \ee
 where $m$ and $n$ are both non-negative integers; $\alpha_i$
 and $\beta_i$ are all constants. Obviously, it reduces to
 the Taylor polynomial when all $\beta_i=0$.

In the present work, we are interested to study the luminosity
 distance and EoS of dark energy by using the Pad\'e approximant. In
 fact, computing the luminosity distance involves repeated
 numerical integral, and hence consumes a large amount of time
 and computing power, especially in the massive computation.
 So, to have a rapid computation, it is desirable to find an
 analytical approximation of the luminosity distance, in place
 of integral. In Sec.~\ref{sec2}, we will discuss this issue
 by the help of Pad\'e approximant. On the other hand, although
 there are many parameterizations for the EoS of dark energy in
 the literature, most of them are {\it ad hoc} and purely
 written by hand. Motivated by the Pad\'e approximant, in
 Sec.~\ref{sec3} we propose two types of Pad\'e
 parameterizations for the EoS of dark energy, and constrain
 them by using the latest observational data. Finally,
 some discussions are given in Sec.~\ref{sec4}.

%============================= section 2 ===================================

\section{Pad\'e analytical approximation of the luminosity
 distance}\label{sec2}

%============================= section 2.1 ===================================

\subsection{Status of the art}\label{sec2a}

As mentioned above, computing the luminosity distance involves
 repeated numerical integral, or elliptic functions~\cite{r8}.
 In order to accelerate a massive computation (e.g. Monte Carlo
 simulation), an analytical approximation of the luminosity
 distance is desirable. To our knowledge, in 1999,
 Pen~\cite{r9} obtained the first analytical approximation
 of the luminosity distance for the flat $\Lambda$CDM model, namely
 \be{eq2}
 d_L=\frac{c}{H_0}(1+z)\left[\eta\left(1,\,\Omega_{m0}\right)-
 \eta\left(\frac{1}{1+z},\,\Omega_{m0}\right)\right]\,,
 \ee
 where
 \bea
 &\disp\eta\left(a,\Omega_{m0}\right)=2\sqrt{s^3+1}\left[
 \frac{1}{a^4}-0.1540\frac{s}{a^3}+0.4304\frac{s^2}{a^2}+
 0.19097\frac{s^3}{a}+0.066941 s^4\right]^{-1/8},\label{eq3}\\
 &\disp s^3=\frac{1-\Omega_{m0}}{\Omega_{m0}}\,,\label{eq4}
 \eea
 and $\Omega_{m0}$ is the present fractional energy density of
 the pressureless matter; $c$ is the speed of light; $H_0$ is
 the Hubble constant; $z$ is the redshift; $a=(1+z)^{-1}$ is
 the scale factor (we have set $a_0=1$; the subscript ``0''
 indicates the present value of corresponding quantity). It is
 claimed that this formula has a relative error of less than
 $0.4\%$ for $0.2<\Omega_{m0}<1$ for any redshift, and a
 global relative error of less than $4\%$ for any choice of
 parameters~\cite{r9}. More than ten years passed, and dark
 energy cosmology has been developed significantly. In the
 recent years, the repeated computation of the luminosity
 distance has become more and more massive, while the observational
 data accumulated significantly. Wickramasinghe
 and Ukwatta~\cite{r10} found a new analytical approximation of the
 luminosity distance for the flat $\Lambda$CDM model, namely
 \be{eq5}
 d_L=\frac{c}{3H_0}\frac{1+z}{\left(1-\Omega_{m0}\right)^{1/6}
 \Omega_{m0}^{1/3}}\left[\Psi(x(0,\,\Omega_{m0}))-\Psi(x(z,\,
 \Omega_{m0}))\right]\,,
 \ee
 where
 \bea
 &\disp\Psi(x)=3x^{1/3}2^{2/3}\left(1-\frac{x^2}{252}+
 \frac{x^4}{21060}\right)\,,\label{eq6}\\[1mm]
 &\disp x(z,\,\Omega_{m0})=\ln\left(\alpha+\sqrt{\alpha^2-1}
 \right)\,,~~~~~~~\alpha(z,\,\Omega_{m0})=
 1+\frac{1-\Omega_{m0}}{\Omega_{m0}}\frac{2}{(1+z)^3}\,.\label{eq7}
 \eea
 They claimed that this formula has a relative error smaller
 than the one of Pen~\cite{r9}. Then, Adachi and
 Kasai~\cite{r6} found another analytical approximation of the
 luminosity distance for the flat $\Lambda$CDM model, namely
 \be{eq8}
 d_L=\frac{2c}{H_0}\frac{1+z}{\sqrt{\Omega_{m0}}}\left[
 \Phi(x(0,\,\Omega_{m0}))-\frac{1}{\sqrt{1+z}}
 \Phi(x(z,\,\Omega_{m0}))\right]\,,
 \ee
 where
 \be{eq9}
 \Phi(x)=\frac{1+1.320x+0.4415x^2+0.02656x^3}{1+1.392x+
 0.5121x^2+0.03944x^3}\,,~~~~~~~x(z,\,\Omega_{m0})=\frac{1-
 \Omega_{m0}}{\Omega_{m0}}\frac{1}{(1+z)^3}\,.
 \ee
 They claimed that for a wide range of $\Omega_{m0}$ and
 redshift $z$, this formula has a relative error even smaller
 than the one of Wickramasinghe and Ukwatta~\cite{r10}. It is
 worth noting that Zhang {\it et al.}~\cite{r11} also discussed
 the computation of the luminosity distance. However, they did
 not obtain the analytical approximation of the luminosity
 distance. Instead, they considered the numerical algorithms
 to compute the elliptic integrals of the luminosity distance.
 So, the works of Zhang {\it et al.}~\cite{r11} are not tightly
 relevant.

%============================= section 2.2 ===================================

\subsection{Analytical approximation of the luminosity distance
 for the flat XCDM model}\label{sec2b}

To our knowledge, the relevant works in the literature concentrated
 on the flat $\Lambda$CDM model, namely the role of dark energy
 is played by a cosmological constant (its EoS $w_{de}=-1$ exactly).
 However, dynamical dark energy with a EoS $w_{de}\not=-1$ is
 extensively considered in cosmology~\cite{r1}. So, it is of
 interest to find also an analytical approximation of the
 luminosity distance for dynamical dark energy models. In the
 present work, we would like to consider a flat XCDM model and
 find an analytical approximation of the luminosity distance in
 this case.

The flat XCDM model describes a flat Friedmann-Robertson-Walker
 (FRW) universe containing only pressureless matter and dark
 energy with $w_{de}=w_X=const.$. To accelerate the cosmic
 expansion, $w_X<0$ is required. By definition, the luminosity
 distance reads (see e.g.~\cite{r6,r9,r10,r12})
 \be{eq10}
 d_L\equiv c\,(1+z)\int_0^z\frac{d\tilde{z}}{H(\tilde{z})}=
 \frac{c}{H_0}(1+z)\int_a^1\frac{d\tilde{a}}{\tilde{a}^2
 E(\tilde{a})}\,,
 \ee
 where $E\equiv H/H_0$ and $H\equiv\dot{a}/a$ is the Hubble
 parameter (a dot denotes the derivative with respect to cosmic
 time $t$). For convenience, we introduce a new function
 \be{eq11}
 \Psi\equiv\frac{1}{2}\int_0^a\frac{\Omega_{m0}^{1/2}\,
 d\tilde{a}}{\tilde{a}^2 E(\tilde{a})}\,.
 \ee

% we insert Figs. 1 and 2 here just for a more comfortable typesetting
% note that Eq. (11) is connected with Eq. (12) in fact

%============================= Fig. 1 =================================

 \begin{center}
 \begin{figure}[tbhp]
 \centering
 \includegraphics[width=0.75\textwidth]{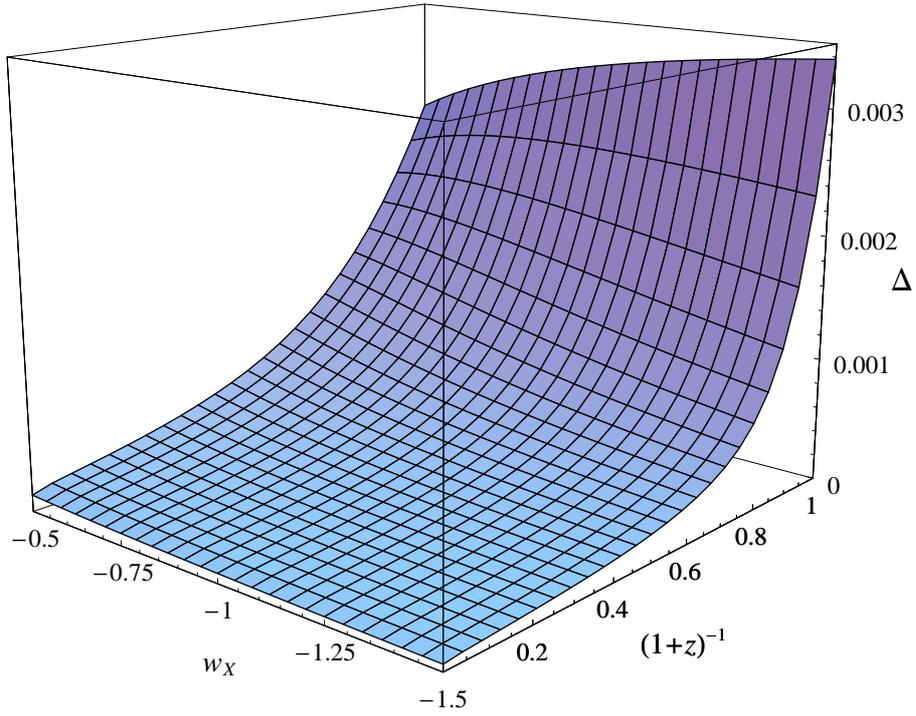}
 \caption{\label{fig1}
 The 3D plot of the relative error $\Delta$ as a function of
 $w_X$ and redshift $z$ while $\Omega_{m0}=0.3$.}
 \end{figure}
 \end{center}

%======================================================================

\vspace{5mm} % used here just for a more comfortable typesetting

%============================= Fig. 2 =================================

 \begin{center}
 \begin{figure}[tbhp]
 \centering
 \includegraphics[width=0.75\textwidth]{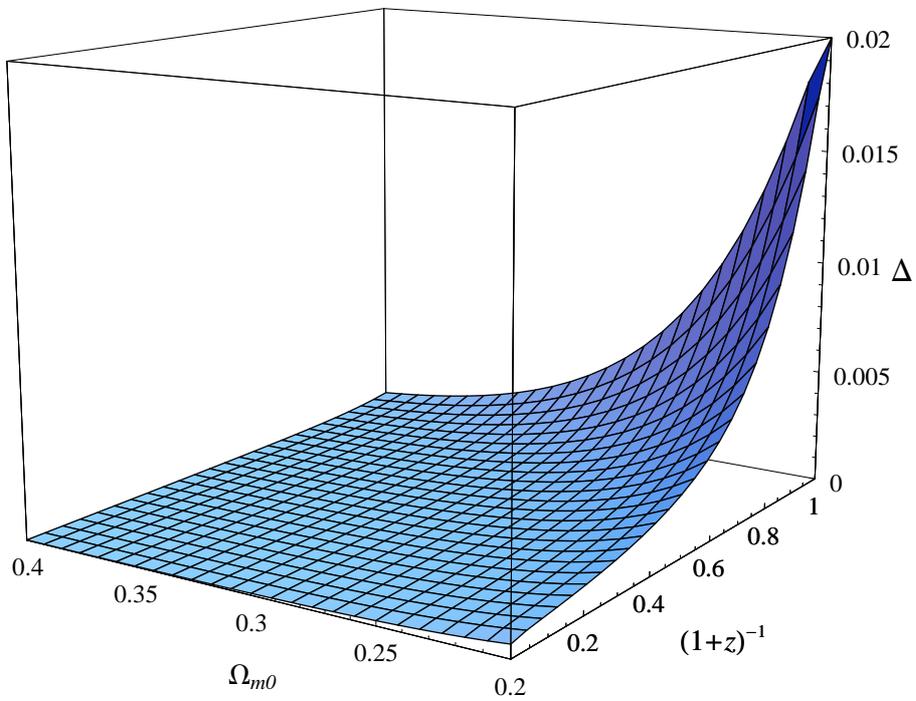}
 \caption{\label{fig2}
 The 3D plot of the relative error $\Delta$ as a function
 of $\Omega_{m0}$ and redshift $z$ while $w_X=-0.95$.}
 \end{figure}
 \end{center}

%======================================================================

\vspace{-22.5mm} % used here just for a more comfortable typesetting

\noindent Using this function, the luminosity distance can be recast as
 \be{eq12}
 d_L=\frac{2c}{H_0}\frac{1+z}{\Omega_{m0}^{1/2}}
 \left[\Psi(a=1)-\Psi(a)\right]\,.
 \ee
 For the flat XCDM model, the corresponding $E=H/H_0$ reads
 (see e.g.~\cite{r12})
 \be{eq13}
 E=\left[\Omega_{m0}\,a^{-3}+\left(1-\Omega_{m0}\right)
 a^{-3(1+w_X)}\right]^{1/2}=\left[\Omega_{m0}(1+z)^3+\left(1-
 \Omega_{m0}\right)(1+z)^{3(1+w_X)}\right]^{1/2}.
 \ee
 Substituting it into Eq.~(\ref{eq11}), we obtain
 \be{eq14}
 \Psi=\frac{1}{2}\int_0^a\frac{1}{\sqrt{\tilde{a}}}\frac{d\tilde{a}}
 {\sqrt{1+s\tilde{a}^{-3w_X}}}=\int_0^x\left(-\frac{1}{6w_X}\right)
 \left(\frac{\tilde{x}}{s}\right)^{-1/(6w_X)}\frac{d\tilde{x}}
 {\tilde{x}\sqrt{1+\tilde{x}}}\,,
 \ee
 where
 \be{eq15}
 x=s\,a^{-3w_X}=s(1+z)^{3w_X}\,,~~~
 ~~~~s\equiv\frac{1-\Omega_{m0}}{\Omega_{m0}}\,.
 \ee
 Considering the Pad\'e approximant of $\Psi(x)$ up to
 order $(3,\,3)$, and noting $a=(x/s)^{-1/(3w_X)}$, we have
 \be{eq16}
 \Psi=\sqrt{a}\cdot\frac{1+\alpha_1 x+\alpha_2 x^2
 +\alpha_3 x^3}{1+\beta_1 x+\beta_2 x^2+\beta_3 x^3}\,,
 \ee
 where the constant coefficients are given by
 \bea
 \alpha_1 &=& 80\,\xi^{-1} (-1 + 12 w_X) (-1 + 18 w_X) (-1 + 24 w_X)
   (-1 + 30 w_X) (1 + 36 w_X (-4 + 3 w_X (107 + \nonumber\\
   & &4 w_X (-1216 + 9 w_X (3805 + 12 w_X (-5888 + 3 w_X (25945 + 48
   w_X (-4469 + 16947 w_X))))))))\,,\nonumber\\
 \alpha_2 &=& 24\,\xi^{-1} (-1 + 18 w_X) (-1 + 24 w_X) (1 +
   36 w_X (-5 + w_X (491 + 36 w_X (-800 + \nonumber \\
   & & 9 w_X (3383 + 4 w_X (-22237 + w_X (427301 + 72 w_X (-82459 +
    8 w_X (97171 + \nonumber\\
   & & 18 w_X (-31315 + 99918 w_X))))))))))\,, \nonumber\\
 \alpha_3 &=& \xi^{-1} (1 + 72 w_X (-3 + 2 w_X (175 + 6 w_X (-2098 +
   3 w_X (33719 + 24 w_X (-48254 +\nonumber\\
   & & 3 w_X (413179 + 3 w_X (-2674436 + 3 w_X (12787417 +
   72 w_X (-1828153 +\nonumber\\
   & & 16 w_X (837203 + 9 w_X (-495569 + 2012094 w_X))))))))))))\,,
   \nonumber\\
 \beta_1 &=& 112\,\xi^{-1} (1 - 18 w_X)^2 (-1 + 6 w_X)
   (-1 + 12 w_X) (-1 + 24 w_X) (-1 + 30 w_X) \nonumber\\
   & &\times (1 + 24 w_X (-5 + 6 w_X (58 + 3 w_X (-688 +
   3 w_X (4687 + 48 w_X (-1243 + 12105 w_X))))))\,,\nonumber\\
 \beta_2 &=& 56\,\xi^{-1} (-1 + 6 w_X) (-1 + 24 w_X) (1 +
   6 w_X (-5 + 36 w_X))^2\nonumber\\
   & &\times (1 + 12 w_X (-11 + 18 w_X (47 + 2 w_X (-925 +
   3 w_X (6989 + 648 w_X (-151 + 1586 w_X))))))\,,\nonumber\\
 \beta_3 &=& 7\,\xi^{-1} (1 - 18 w_X)^2 (1 - 12 w_X)^2 (1 - 6 w_X)^2
   \nonumber\\
   & &\times (1 + 72 w_X (-2 + 3 w_X (57 + 2 w_X (-1252 +
   9 w_X (3647 + 144 w_X (-421 + 5323 w_X))))))\,,\label{eq17}
 \eea
 in which
 \bea
 \xi&\equiv& 64 (-1 + 6 w_X) (-1 + 12 w_X) (-1 + 18 w_X)
   (-1 + 24 w_X) (-1 + 30 w_X) (-1 + 36 w_X)\nonumber\\
   & &\times (1 + 108 w_X (-1 + 4 w_X (16 + w_X (-521 +
   9 w_X (1105 + 24 w_X (-555 + 5186 w_X))))))\,.\label{eq18}
 \eea
 Although the constant coefficients in Eqs.~(\ref{eq17}) and
 (\ref{eq18}) look awesome, it is not a problem when one
 calculate the luminosity distance using computer (see the
 discussions in Sec.~\ref{sec4}). It is easy to check that if
 $w_X=-1$, our results can reduce to the one of Adachi and
 Kasai~\cite{r6}, namely Eqs.~(\ref{eq8}) and (\ref{eq9}).

\begin{center}
 \begin{figure}[tbhp]
 \centering
 \includegraphics[width=1.0\textwidth]{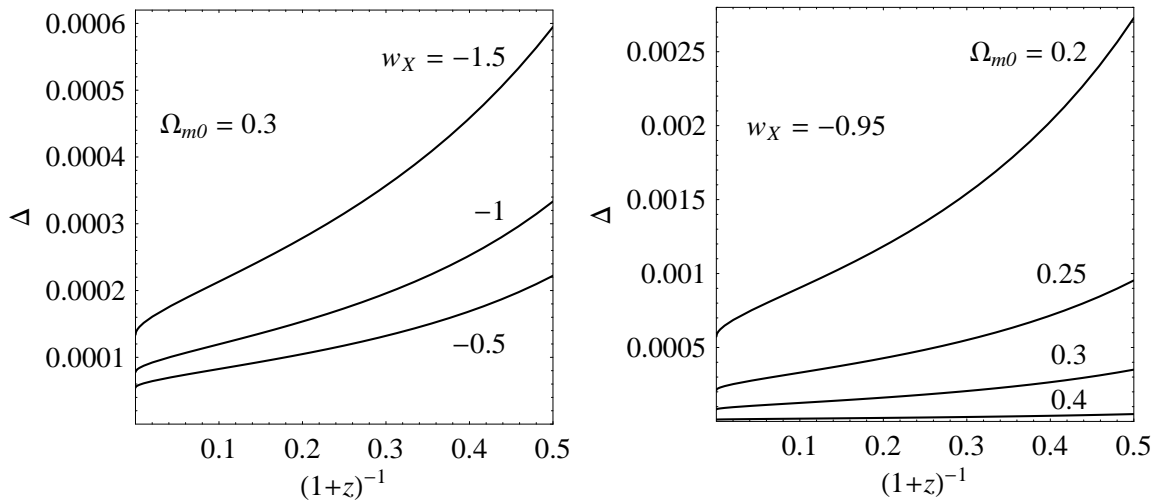}
 \caption{\label{fig3}
 The 2D plot of the relative error $\Delta$ for redshift $z\geq 1$.
 In the left panel, $\Omega_{m0}=0.3$ is fixed, while the lines
 from top to bottom correspond to $w_X=-1.5$, $-1$, $-0.5$,
 respectively. In the right panel, $w_X=-0.95$ is fixed, while
 the lines from top to bottom correspond to $\Omega_{m0}=0.2$,
 $0.25$, $0.3$, $0.4$, respectively.}
 \end{figure}
 \end{center}

Since our analytical approximation of the luminosity distance
 for the flat XCDM model is obtained for the first time (to
 our knowledge), we can only compare it with the exact integral
 in Eq.~(\ref{eq10}). The relative error reads
 \be{eq19}
 \Delta\equiv\frac{d_L^{Pade}-d_L^{int}}{d_L^{int}}\,,
 \ee
 where $d_L^{Pade}$ is calculated using Eq.~(\ref{eq12}), while
 $\Psi$ is given in Eq.~(\ref{eq16}); $d_L^{int}$ is calculated
 using Eq.~(\ref{eq10}), while $E$ is given in
 Eq.~(\ref{eq13}). We present the 3D plots of the relative
 error $\Delta$ as a function of $w_X$, $\Omega_{m0}$ and
 redshift $z$ in Figs.~\ref{fig1} and \ref{fig2}.
 In Fig.~\ref{fig1}, we fixed $\Omega_{m0}=0.3$. In the whole
 history $0\leq z<\infty$, for any EoS of dark
 energy in the range $-1.5\leq w_X\leq -0.5$, the relative
 error $\Delta$ is always smaller than $0.34\%$. The larger
 $w_X$, the smaller relative error $\Delta$ is. It is easy to
 see that the relative error $\Delta$ is not sensitive to the
 EoS of dark energy $w_X$ in fact. The relative error $\Delta$
 increases only when $z\,\lsim\,1$. To see clearly, in
 Fig.~\ref{fig3}, we present the 2D plots of $\Delta$ for
 $z\geq 1$. From the left panel of Fig.~\ref{fig3}, for a fixed
 $\Omega_{m0}=0.3$ and any EoS of dark energy in the range
 $-1.5\leq w_X\leq -0.5$, the relative error $\Delta$ is always
 smaller than $0.06\%$ for redshift $z\geq 1$. For $w_X$ around
 $-1$, the relative error $\Delta$ is always smaller than $0.03\%$.
 For high redshift, the relative error $\Delta$ is smaller than
 $0.008\%$. On the other hand, in Fig.~\ref{fig2}, we fixed
 $w_X=-0.95$ instead. The relative error $\Delta$ is sensitive
 to $\Omega_{m0}$ when it is smaller than $0.25$. In the whole
 history $0\leq z<\infty$, for any $\Omega_{m0}$ in the range
 $0.2\leq\Omega_{m0}\leq 0.4$, the relative error $\Delta$ is
 always smaller than $2\%$. The smaller $\Omega_{m0}$, the
 larger relative error $\Delta$ is. Fortunately, the latest
 Planck 2013 data~\cite{r13} favors a large $\Omega_{m0}\sim 0.315$.
 For this large $\Omega_{m0}$, the relative error $\Delta$ is
 always smaller than $0.3\%$ in the whole history $0\leq z<\infty$.
 From the right panel of Fig.~\ref{fig3}, for a
 fixed $w_X=-0.95$ and any $\Omega_{m0}$ in the range
 $0.2\leq\Omega_{m0}\leq 0.4$, the relative error $\Delta$ is
 always smaller than $0.27\%$ for redshift $z\geq 1$. For
 $\Omega_{m0}$ around $0.3$, the relative error $\Delta$ is
 always smaller than $0.03\%$. For high redshift, the relative
 error $\Delta$ is smaller than $0.02\%$.

%============================= Fig. 3 =================================

%======================================================================

\vspace{-9mm} % used here just for a more comfortable typesetting

%============================= section 3 ===================================

\section{Pad\'e parameterizations for the EoS
 of dark energy}\label{sec3}

%============================= section 3.1 ===================================

\subsection{Various EoS parameterizations in
 the literature}\label{sec3a}

Now, we turn to another issue, namely the parameterizations for
 the EoS of dark energy. Today, there are many EoS parameterizations
 in the literature. In the early researches, the popular
 parameterization is given by~\cite{r14}
 \be{eq20}
 w_{de}=w_0+w_1 z\,.
 \ee
 Its generalized parameterization is
 $w_{de}=w_0+w_1 z+w_2 z^2$~\cite{r15}. They can be regarded as
 the Taylor series expansion of $w_{de}$ with respect to
 redshift $z$ up to first or second order. However, these two
 parameterizations cannot work well when redshift $z$ is
 high. So, they have been soon replaced by the well-known
 Chevallier-Polarski-Linder (CPL) parameterization~\cite{r16}
 \be{eq21}
 w_{de}=w_0+w_a(1-a)=w_0+w_a\frac{z}{1+z}\,.
 \ee
 Its generalized parameterization is
 $w_{de}=w_0+w_a(1-a)+w_b(1-a)^2$~\cite{r17}. They can be
 regarded as the Taylor series expansion of $w_{de}$ with
 respect to $(1-a)$ (or scale factor $a$) up to first or second
 order. In the passed ten years, the CPL parameterization is
 the most popular one and has been extensively used in the
 literature. Although the CPL parameterization dominated most
 works, there are still many exotic parameterizations in the
 literature. For instance, in e.g.~\cite{r18}, the following
 parameterization has been proposed, namely
 \be{eq22}
 w_{de}=w_0+w_1\frac{z}{(1+z)^\alpha}\,,
 \ee
 where $\alpha$ usually was taken to be 2. An interesting
 parameterization was considered in e.g.~\cite{r19}, i.e.,
 \be{eq23}
 w_{de}=w_0+w_a\frac{1-a^\beta}{\beta}\,.
 \ee
 The logarithm parameterization~\cite{r20},
 \be{eq24}
 w_{de}=w_0+w_1\ln a\,,
 \ee
 can be seen in many works in the literature. Another logarithm
 parameterization reads~\cite{r21}
 \be{eq25}
 w_{de}=\frac{w_0}{\left[1+b\ln(1+z)\right]^\alpha}\,,
 \ee
 where $\alpha$ usually was taken to be 1 or 2.

Although the CPL parameterization works well, it will diverge
 when $a\to\infty$ (or equivalently $z\to -1$). This is also a
 common feature of many existing parameterizations. To overcome
 this divergence, several {\it ad hoc} parameterizations have
 been proposed in the literature. For example, almost six years
 ago the following parameterization~\cite{r22} has been
 proposed to avoid the divergence at $z\to -1$, namely
 \be{eq26}
 w_{de}=w_0+w_1\frac{z(1+z)}{1+z^2}\,.
 \ee
 Only one month later, a similar parameterization~\cite{r23}
 has been proposed, i.e.,
 \be{eq27}
 w_{de}=\frac{w_0}{1+\left(w_1 z\right)^2}\,.
 \ee
 Recently, this issue regained attention. In~\cite{r24}, the
 following parameterization was proposed to avoid the divergence at
 $z\to -1$, namely
 \be{eq28}
 w_{de}=w_0+w_1\left[\frac{\ln (2+z)}{1+z}-\ln 2\right]\,.
 \ee
 Two parameterizations similar to the one in Eq.~(\ref{eq26})
 were considered in~\cite{r25}, i.e.,
 \be{eq29}
 w_{de}=w_0+w_1\frac{z}{1+z^2}\,,~~~~~~~~
 w_{de}=w_0+w_1\frac{z^2}{1+z^2}\,.
 \ee
 Actually, in the literature, there are other {\it ad hoc}
 parameterizations to this end. We finally mention two of
 them~\cite{r26}, namely
 \be{eq30}
 w_{de}=w_0+w_1\frac{z}{2+z}\,,~~~~~~
 w_{de}=w_0+w_1\frac{1-a}{1+a}\,.
 \ee
 We refer to e.g.~\cite{r1,r27} and the references therein for
 other exotic parameterizations.

%============================= section 3.2 ===================================

\subsection{Type (I) Pad\'e parameterization}\label{sec3b}

As mentioned above, most of the existing EoS parameterizations
 are {\it ad hoc} and purely written by hand. In particular,
 all the parameterizations without the divergence at $z\to -1$
 ($a\to\infty$) are not well motivated from mathematics or
 fundamental physics. So, the well-motivated parameterizations
 are still welcome. As mentioned in Sec.~\ref{sec1}, in
 mathematics, any function could be approximated by the
 Pad\'e approximant. The Pad\'e approximant is the best
 approximation of a function by a rational function of given
 order~\cite{r4}. In fact, the Pad\'e approximant often gives
 better approximation of the function than truncating its
 Taylor series, and it may still work where the Taylor series
 does not converge~\cite{r4}. This fact motivates us to propose
 several novel EoS parameterizations here.

Firstly, we consider the type (I) Pad\'e parameterization,
 \be{eq31}
 w_{de}=\frac{w_0+w_a(1-a)}{1+w_b(1-a)}\,,
 \ee
 where $w_0$, $w_a$ and $w_b$ are all constants. In fact, it is
 the Pad\'e approximant of $w_{de}$ with respect to $(1-a)$ (or
 scale factor $a$) up to order $(1,\,1)$. Noting that it can
 be recast as
 \be{eq32}
 w_{de}=\frac{w_0+(w_0+w_a)z}{1+(1+w_b)z}\,,
 \ee
 it is also the Pad\'e approximant of $w_{de}$ with respect
 to redshift $z$ up to order $(1,\,1)$. It is worth noting
 that if $w_b=0$, our type (I) Pad\'e parameterization
 (\ref{eq31}) or (\ref{eq32}) can reduce to the well-known
 CPL parameterization~(\ref{eq21}). If $w_b\not=0$, our
 type (I) Pad\'e parameterization can avoid the divergence at
 $a\to\infty$ (or $z\to -1$ equivalently), unlike the CPL
 parameterization. In fact, it is easy to see that
 \be{eq33}
 w_{de}=\left\{
 \begin{array}{ll}
  \disp\frac{w_0+w_a}{1+w_b}\,,~~ & {\rm for}~~
        a\to 0 ~(z\to\infty,~{\rm the~early~time})\,,\\[5mm]
     w_0\,, & {\rm for}~~ a=1 ~(z=0,~{\rm now})\,,\\[4mm]
     \disp\frac{w_a}{w_b}\,, & {\rm for}~~ a\to\infty~
          (z\to -1,~{\rm the~far~future})\,,
  \end{array} \right.
 \ee
 where $w_b\not=0$ and $w_b\not=-1$ are required. In addition,
 to avoid the denominator in Eq.~(\ref{eq31}) being zero
 for any physical scale factor $a\geq 0$, we require that
 \be{eq34}
 -1<w_b<0\,.
 \ee
 So, the denominator in Eq.~(\ref{eq31}) is always positive.
 Under the condition in Eq.~(\ref{eq34}), our type (I) Pad\'e
 parameterization is always regular for the whole
 $0\leq a<\infty$ (or $-1\leq z<\infty$ equivalently).
 Note that $w_{de}$ can cross the so-called phantom divide
 $w_{de}=-1$ at
 \be{eq35}
 a_\ast=1+\frac{1+w_0}{w_a+w_b}\,,~~~~~~~{\rm or~equivalently,}
 ~~~~~~~z_\ast=\frac{-1-w_0}{1+w_0+w_a+w_b}\,.
 \ee

Naturally, it is important to confront our Pad\'e
 parameterization with the latest observational data. The
 Union2.1 compilation~\cite{r2} is the largest published and
 spectroscopically confirmed Type Ia supernovae (SNIa) sample
 to date. The 580 data points of Union2.1 SNIa compilation are
 given in terms of the distance modulus $\mu_{obs}(z_i)$. On
 the other hand, the theoretical distance modulus is defined by
 \be{eq36}
 \mu_{th}(z_i)\equiv 5\log_{10}D_L(z_i)+\mu_0\,,
 \ee
 where $\mu_0\equiv 42.38-5\log_{10}h$ and $h$ is the Hubble
 constant $H_0$ in units of $100~{\rm km/s/Mpc}$, while
 \be{eq37}
 D_L(z)=(1+z)\int_0^z
 \frac{d\tilde{z}}{E(\tilde{z};{\bf p})}\,,
 \ee
 in which ${\bf p}$ denotes the model parameters,
 and $E\equiv H/H_0$. Correspondingly, the $\chi^2$ from 580
 Union2.1 SNIa is given by
 \be{eq38}
 \chi^2_{\mu}({\bf p})=
 \sum\limits_{i}\frac{\left[\mu_{obs}(z_i)-\mu_{th}(z_i)
 \right]^2}{\sigma^2(z_i)}\,,
 \ee
 where $\sigma$ is the corresponding $1\sigma$
 error. The parameter $\mu_0$ is a nuisance
 parameter but it is independent of the data points. One can
 perform a uniform marginalization over $\mu_0$. However,
 there is an alternative way. Following~\cite{r28,r29}, the
 minimization with respect to $\mu_0$ can be made
 by expanding the $\chi^2_{\mu}$ of Eq.~(\ref{eq38}) with
 respect to $\mu_0$ as
 \be{eq39}
 \chi^2_{\mu}({\bf p})=
 \tilde{A}-2\mu_0\tilde{B}+\mu_0^2\tilde{C}\,,
 \ee
 where
 \bea
 &\disp\tilde{A}({\bf p})=\sum\limits_{i}\frac{\left[\mu_{obs}(z_i)
 -\mu_{th}(z_i;\mu_0=0,{\bf p})\right]^2}
 {\sigma_{\mu_{obs}}^2(z_i)}\,,\nonumber \\
 &\disp\tilde{B}({\bf p})=\sum\limits_{i}\frac{\mu_{obs}(z_i)-
 \mu_{th}(z_i;\mu_0=0,{\bf p})}{\sigma_{\mu_{obs}}^2(z_i)}\,,
 ~~~~~~~~~~\tilde{C}=\sum\limits_{i}
 \frac{1}{\sigma_{\mu_{obs}}^2(z_i)}\,.\nonumber
 \eea
 Eq.~(\ref{eq39}) has a minimum for
 $\mu_0=\tilde{B}/\tilde{C}$ at
 \be{eq40}
 \tilde{\chi}^2_{\mu}({\bf p})=
 \tilde{A}({\bf p})-\frac{\tilde{B}({\bf p})^2}{\tilde{C}}\,.
 \ee
 Since $\chi^2_{\mu,\,min}=\tilde{\chi}^2_{\mu,\,min}$
 (up to a constant) obviously, we can instead minimize
 $\tilde{\chi}^2_{\mu}$ which is independent of $\mu_0$.
 In addition to SNIa, the other useful observations include
 the cosmic microwave background (CMB) anisotropy~\cite{r13}
 and the large-scale structure (LSS)~\cite{r17,r30}. However,
 using the full data of CMB and LSS to perform a global
 fitting consumes a large amount of computation time and
 power. As an alternative, one can instead use the shift
 parameter $R$ from CMB, and the distance parameter $A$ of
 the measurement of the baryon acoustic oscillation (BAO)
 peak in the distribution of SDSS luminous red galaxies. In
 the literature, the shift parameter $R$ and the distance
 parameter $A$ have been used extensively. It is argued in
 e.g.~\cite{r31} that they are model-independent and contain
 the main information of the observations of CMB and BAO,
 respectively. As is well known, the shift parameter $R$ of
 CMB is defined by~\cite{r31,r32}
 \be{eq41}
 R\equiv\Omega_{m0}^{1/2}\int_0^{z_{rec}}
 \frac{d\tilde{z}}{E(\tilde{z})}\,,
 \ee
 where the redshift of recombination $z_{rec}=1090.48$ which
 was determined by the latest Planck 2013 data~\cite{r13},
 and $\Omega_{m0}\equiv 8\pi G\rho_{m0}/(3H_0^2)$ is the
 present fractional density of pressureless matter. The value
 of $R$ has been determined to be $1.7407\pm 0.0094$ from the
 Planck 2013 data~\cite{r33} (note that in the original Planck
 paper~\cite{r13} the value of $R$ has not been given directly,
 and it was obtained in~\cite{r33} later by other authors
 using the Planck 2013 data~\cite{r13}). On the other hand,
 the distance parameter $A$ of the measurement of the BAO
 peak in the distribution of SDSS luminous red
 galaxies~\cite{r17,r30} is given by
 \be{eq42}
 A\equiv\Omega_{m0}^{1/2}E(z_b)^{-1/3}\left[
 \frac{1}{z_b}\int_0^{z_b}\frac{d\tilde{z}}{E(\tilde{z})}
 \right]^{2/3},
 \ee
 where $z_b=0.35$. In~\cite{r34}, the value
 of $A$ has been determined to be
 $0.469\,(n_s/0.98)^{-0.35}\pm 0.017$. Here the scalar spectral
 index $n_s$ is taken to be $0.9662$, which comes from the
 Planck 2013 data~\cite{r33}. So, the total $\chi^2$ is
 given by
 \be{eq43}
 \chi^2=\tilde{\chi}^2_{\mu}+\chi^2_{CMB}+\chi^2_{BAO}\,,
 \ee
 where $\tilde{\chi}^2_{\mu}$ is
 given in Eq.~(\ref{eq40}), $\chi^2_{CMB}=(R-R_{obs})^2/\sigma_R^2$
 and $\chi^2_{BAO}=(A-A_{obs})^2/\sigma_A^2$. The best-fit
 model parameters are determined by minimizing the total
 $\chi^2$. As in~\cite{r28,r12}, the $68.3\%$ confidence
 level is determined by
 $\Delta\chi^2\equiv\chi^2-\chi^2_{min}\leq 1.0$, $2.3$,
 $3.53$, $4.72$ for $n_p=1$, $2$, $3$, $4$ respectively, where
 $n_p$ is the number of free model parameters. Similarly, the
 $95.4\%$ confidence level is determined by
 $\Delta\chi^2\equiv\chi^2-\chi^2_{min}\leq 4.0$, $6.18$,
 $8.02$, $9.72$ for $n_p=1$, $2$, $3$, $4$, respectively.

%============================= Fig. 4 =================================

 \begin{center}
 \begin{figure}[tbhp]
 \centering
 \includegraphics[width=0.849\textwidth]{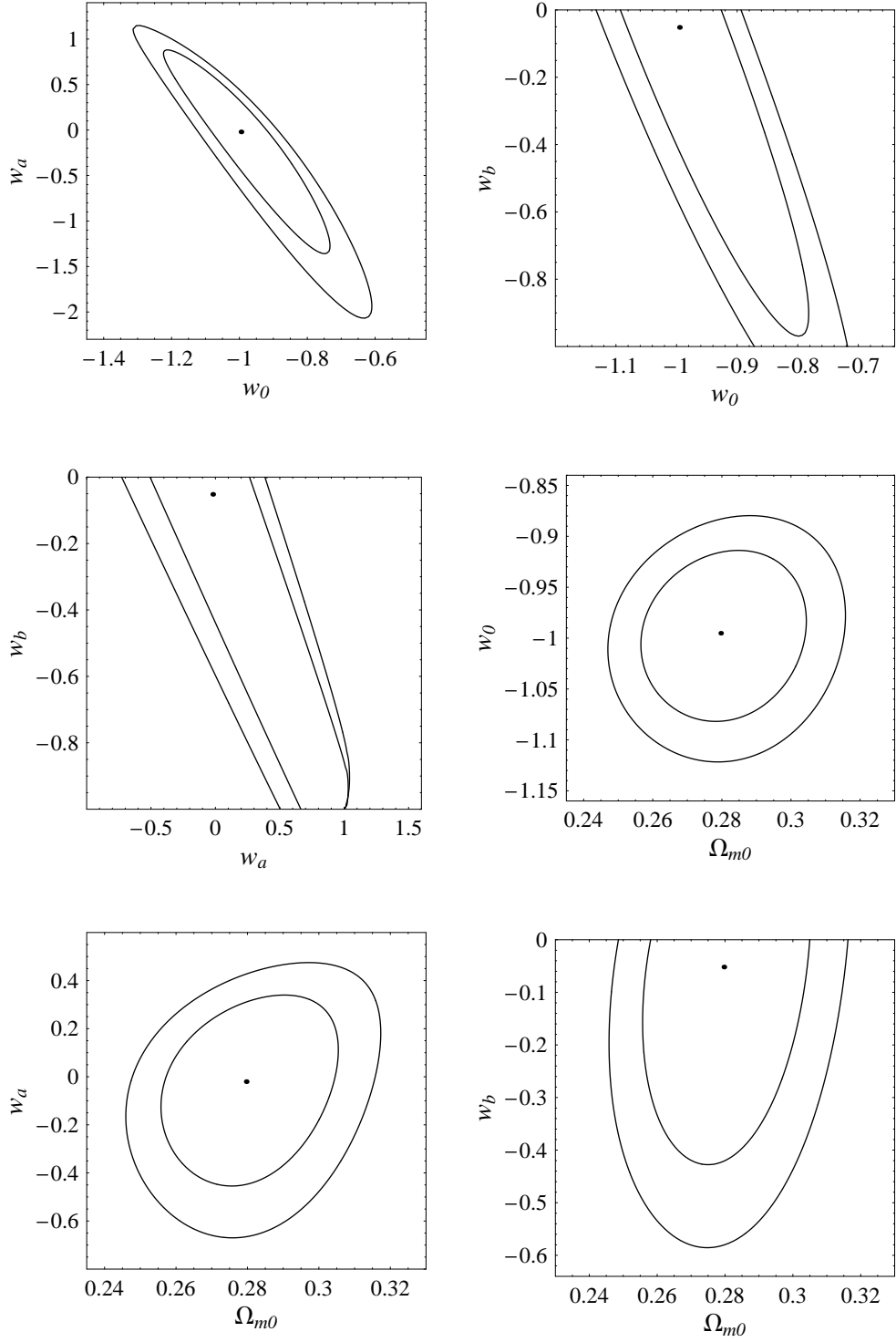}
 \caption{\label{fig4}
 The $68.3\%$ and $95.4\%$ confidence level contours in
 the $w_0-w_a$, $w_0-w_b$, $w_a-w_b$, $\Omega_{m0}-w_0$,
 $\Omega_{m0}-w_a$, and $\Omega_{m0}-w_b$ planes for the
 type (I) Pad\'e parameterization in Eq.~(\ref{eq31}).
 The best-fit parameters are also indicated by the black
 solid points.}
 \end{figure}
 \end{center}

%======================================================================

\vspace{-7mm} % used here just for a more comfortable typesetting

\begin{center}
 \begin{figure}[tbhp]
 \centering
 \includegraphics[width=1.0\textwidth]{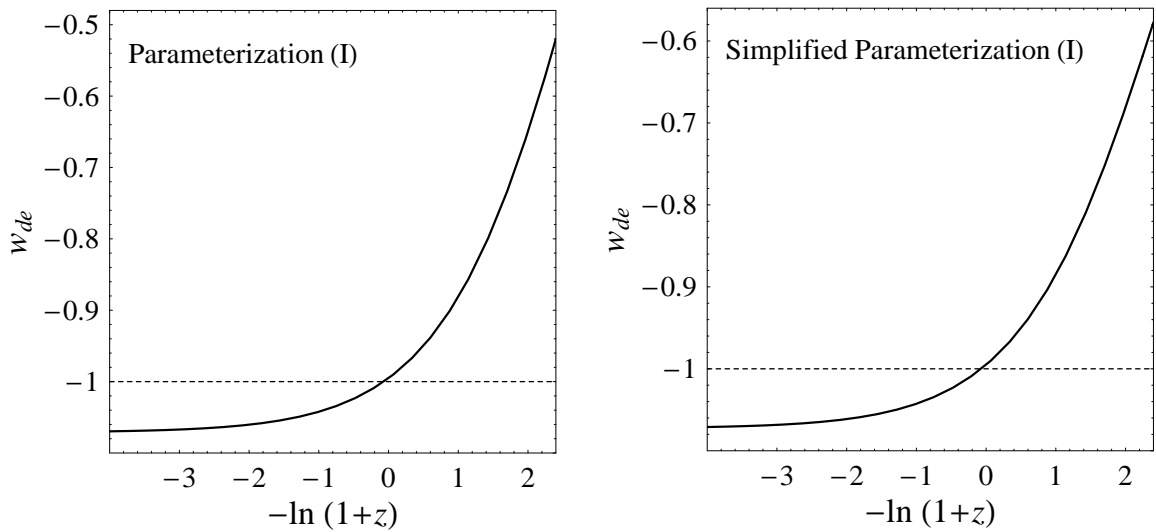}
 \caption{\label{fig5}
 $w_{de}$ as a function of redshift $z$ for the models (left
 panel: parameterization (I) in Eq.~(\ref{eq31}); right panel:
 simplified parameterization (I) in Eq.~(\ref{eq46})) with
 their best-fit parameters, respectively. See the text for
 details.}
 \end{figure}
 \end{center}

Now, let us come back to our type (I) Pad\'e parameterization,
 namely Eq.~(\ref{eq31}). Substituting this $w_{de}$ into the
 energy conservation equation
 $\dot{\rho}_{de}+3H\rho_{de}\left(1+w_{de}\right)=0$, we find that
 \be{eq44}
 \rho_{de}=\rho_{de,0}\,a^{-3(1+w_0+w_a+w_b)/(1+w_b)}\left[
 1+w_b (1-a)\right]^{-3(w_a-w_0 w_b)/[w_b (1+w_b)]}\,.
 \ee
 Note that $w_b\not=0$ and $w_b\not=-1$ as mentioned above. In
 this work, we consider a flat FRW universe containing only
 pressureless matter and dark energy. Substituting Eq.~(\ref{eq44})
 into Friedmann equation, we finally obtain
 \be{eq45}
 E^2=\Omega_{m0}(1+z)^3+\left(1-\Omega_{m0}\right)
 (1+z)^{3(1+w_0+w_a+w_b)/(1+w_b)}\left(1+\frac{w_b\,z}{1+z}
 \right)^{-3(w_a-w_0 w_b)/[w_b (1+w_b)]}\,.
 \ee
 There are four free parameters in this model,
 namely $\Omega_{m0}$, $w_0$, $w_a$ and $w_b$\,. Note that
 $-1<w_b<0$~is required in Eq.~(\ref{eq34}). By minimizing
 the corresponding total $\chi^2$ in Eq.~(\ref{eq43}), we find
 the best-fit parameters $\Omega_{m0}=0.280$, $w_0=-0.995$,
 $w_a=-0.020$, and $w_b=-0.052$, while $\chi^2_{min}=562.256$.
 In Fig.~\ref{fig4}, we present the $68.3\%$ and $95.4\%$
 confidence level contours in the $w_0-w_a$, $w_0-w_b$,
 $w_a-w_b$, $\Omega_{m0}-w_0$, $\Omega_{m0}-w_a$,
 and $\Omega_{m0}-w_b$ planes for the type
 (I) Pad\'e parameterization in Eq.~(\ref{eq31}). Noting
 Eqs.~(\ref{eq33}) and (\ref{eq35}), for the
 best-fit parameters, $w_{de}=-1.071$ in the very beginning
 ($a\to 0$, $z\to\infty$), and then crossed the phantom divide
 at $a_\ast=0.933$ ($z_\ast=0.071$); today ($a=1$, $z=0$), we
 have $w_{de}=-0.995$; in the far future ($a\to\infty$,
 $z\to -1$), $w_{de}=0.393$. We plot this $w_{de}$ as a
 function of redshift $z$ in the left panel of Fig.~\ref{fig5},
 and we can easily see that $w_{de}$ crossed the phantom
 divide $w_{de}=-1$. Obviously, the type (I)
 Pad\'e parameterization works very well in fact.

%============================= Fig. 5 =================================

%======================================================================

\vspace{-7mm} % used here just for a more comfortable typesetting

%============================= Fig. 6 =================================

 \begin{center}
 \begin{figure}[tbhp]
 \centering
 \includegraphics[width=0.849\textwidth]{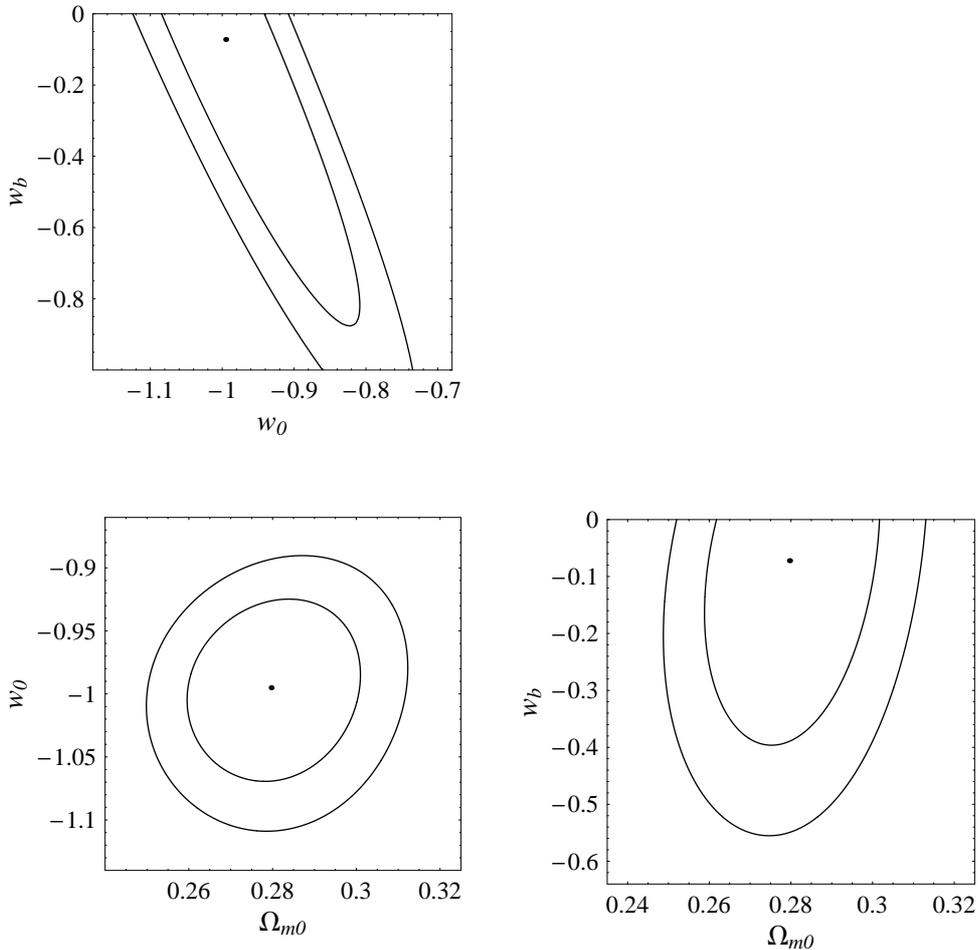}
 \caption{\label{fig6}
 The $68.3\%$ and $95.4\%$ confidence level contours in
 the $w_0-w_b$, $\Omega_{m0}-w_0$, and $\Omega_{m0}-w_b$
 planes for the simplified parameterization (I)
 in Eq.~(\ref{eq46}). The best-fit parameters are
 also indicated by the black solid points.}
 \end{figure}
 \end{center}

%======================================================================

\vspace{-7mm} % used here just for a more comfortable typesetting

Note that the Pad\'e parameterization (I) in Eq.~(\ref{eq31})
 has three free parameters, namely $w_0$, $w_a$ and $w_b$\,. We
 can simplify it by setting $w_a=0$.
 The simplified parameterization (I) reads
 \be{eq46}
 w_{de}=\frac{w_0}{1+w_b (1-a)}\,,
 \ee
 which has only two free parameters $w_0$ and $w_b$\,. It is
 in fact the Pad\'e approximant of $w_{de}$ with respect to $(1-a)$
 (or scale factor $a$) up to order $(0,\,1)$. Of course, the
 condition $-1<w_b<0$ in Eq.~(\ref{eq34}) is still required
 to avoid any singularity. In this case, Eqs.~(\ref{eq33}),
 (\ref{eq35}) and (\ref{eq44}), (\ref{eq45}) are still valid
 but one should set $w_a=0$ in them. By minimizing the
 corresponding total $\chi^2$ in Eq.~(\ref{eq43}), we find
 the best-fit parameters $\Omega_{m0}=0.280$, $w_0=-0.995$, and
 $w_b=-0.072$, while $\chi^2_{min}=562.256$. In Fig.~\ref{fig6}, we
 present the $68.3\%$ and $95.4\%$ confidence level contours in the
 $w_0-w_b$, $\Omega_{m0}-w_0$, and $\Omega_{m0}-w_b$ planes
 for the simplified parameterization (I) in Eq.~(\ref{eq46}).
 For the best-fit parameters, $w_{de}=-1.073$ in the very
 beginning ($a\to 0$, $z\to\infty$), and then crossed the
 phantom divide at $a_\ast=0.932$ ($z_\ast=0.073$); today
 ($a=1$, $z=0$), we have $w_{de}=-0.995$; in the far future
 ($a\to\infty$, $z\to -1$), $w_{de}\to 0$. We plot this
 $w_{de}$ as a function of redshift $z$ in the right panel of
 Fig.~\ref{fig5}, and we can easily see that $w_{de}$ crossed
 the phantom divide $w_{de}=-1$.

%============================= section 3.3 ===================================

\subsection{Type (II) Pad\'e parameterization}\label{sec3c}

\begin{center}
 \begin{figure}[tbhp]
 \centering
 \includegraphics[width=0.849\textwidth]{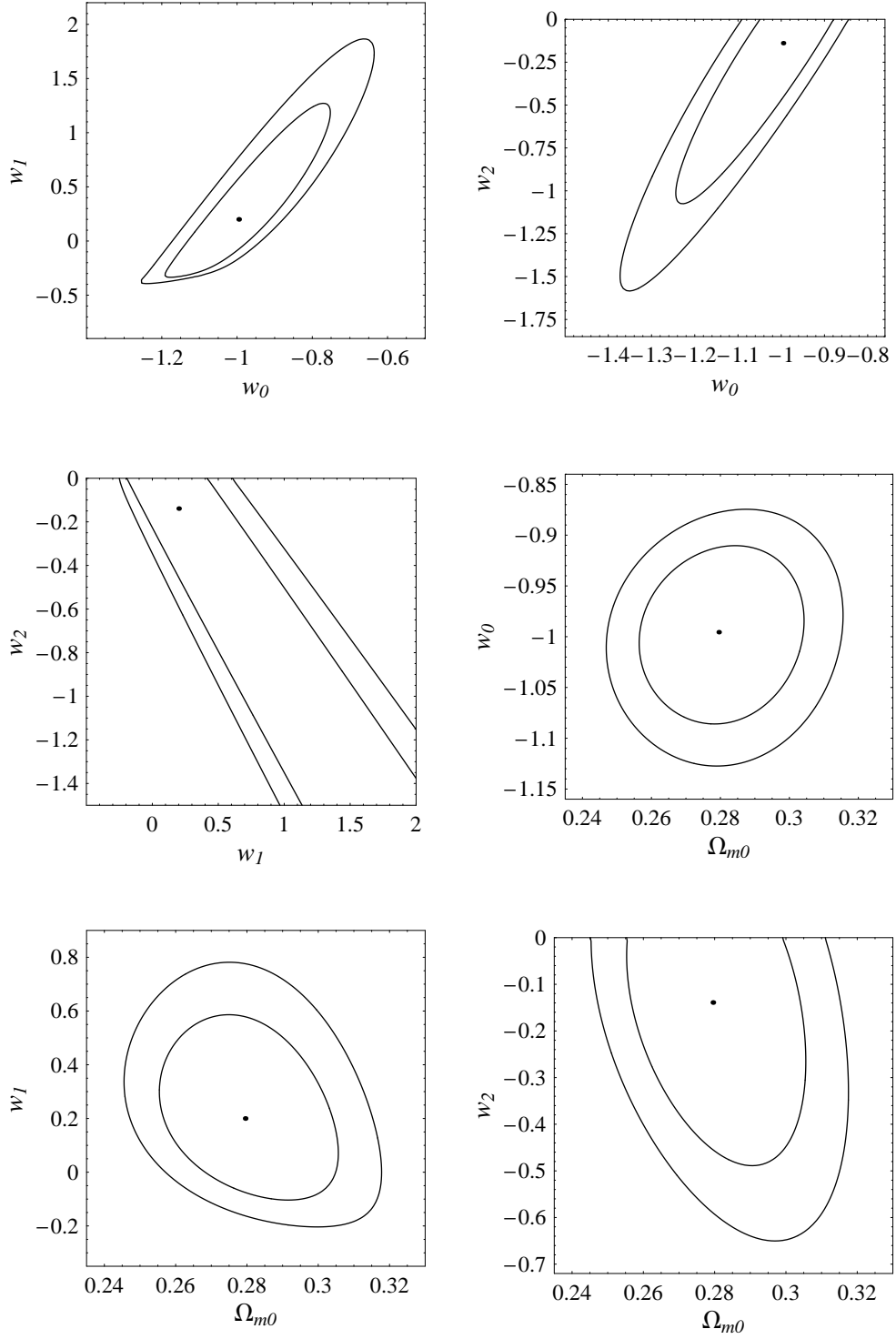}
 \caption{\label{fig7}
 The $68.3\%$ and $95.4\%$ confidence level contours in
 the $w_0-w_1$, $w_0-w_2$, $w_1-w_2$, $\Omega_{m0}-w_0$,
 $\Omega_{m0}-w_1$, and $\Omega_{m0}-w_2$ planes for the
 type (II) Pad\'e parameterization in Eq.~(\ref{eq47}).
 The best-fit parameters are also indicated by the black
 solid points.}
 \end{figure}
 \end{center}

\begin{center}
 \begin{figure}[tbhp]
 \centering
 \includegraphics[width=0.46\textwidth]{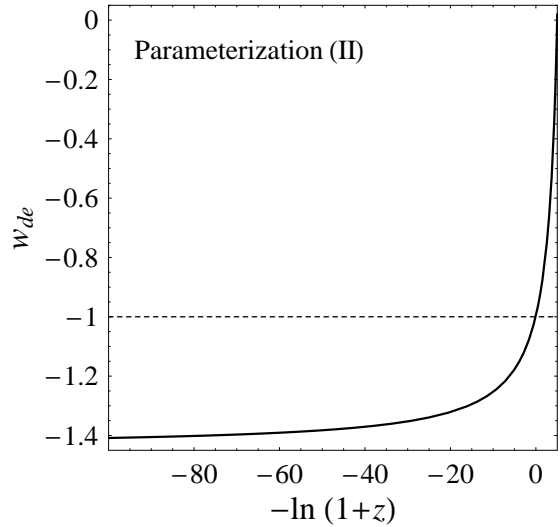}
 \caption{\label{fig8}
 $w_{de}$ as a function of redshift $z$ for the
 parameterization (II) in Eq.~(\ref{eq47})
 with the best-fit parameters. See the text for details.}
 \end{figure}
 \end{center}

In the literature, the most familiar time variables are cosmic
 time $t$, scale factor $a$, redshift $z$, and the so-called
 $e$-folding time $N=\ln a$. As is well known, in many cases
 it is more convenient to express cosmological quantities in
 terms of the $e$-folding time $N=\ln a$. Therefore, it is
 reasonable to consider the Pad\'e approximant of $w_{de}$
 with respect to the $e$-folding time $N=\ln a$ up to order
 $(1,\,1)$, and propose the type (II) Pad\'e parameterization
 \be{eq47}
 w_{de}=\frac{w_0+w_1\ln a}{1+w_2\ln a}\,,
 \ee
 where $w_0$, $w_1$ and $w_2$ are all constants. Obviously,
 it reduces to the parameterization in Eq.~(\ref{eq24}) if
 $w_2=0$, and reduces to the parameterization in
 Eq.~(\ref{eq25}) with $\alpha=1$ if $w_1=0$. If $w_2\not=0$,
 our type (II) Pad\'e parameterization can avoid
 the divergences at $a\to\infty$ (or $z\to -1$ equivalently),
 and $a\to 0$ (or $z\to\infty$ equivalently), unlike the
 logarithm parameterization in Eq.~(\ref{eq24}). It is easy to
 see that
 \be{eq48}
 w_{de}=\left\{
 \begin{array}{ll}
  \disp\frac{w_1}{w_2}\,,~~ & {\rm for}~~
        a\to 0 ~(z\to\infty,~{\rm the~early~time})\,,\\[5mm]
     w_0\,, & {\rm for}~~ a=1 ~(z=0,~{\rm now})\,,\\[4mm]
     \disp\frac{w_1}{w_2}\,, & {\rm for}~~ a\to\infty~
          (z\to -1,~{\rm the~far~future})\,,
  \end{array} \right.
 \ee
 where $w_2\not=0$ is required. However, this parameterization
 unfortunately has an unavoidable singularity at
 $a=\exp (-1/w_2)$ (or $z=-1+\exp (1/w_2)$ equivalently). It
 is a so-called $w$-singularity~\cite{r35} (see below) in
 fact. This type of singularity also exists in the well-known
 logarithm parameterization in Eq.~(\ref{eq25}).
 Nevertheless, if we require
 \be{eq49}
 w_2<0\,,
 \ee
 the singularity will occur in the future, namely $a>1$
 ($z<0$). Under the condition~(\ref{eq49}), the type (II)
 Pad\'e parameterization works well at least in the whole
 past history $0\leq a\leq 1$ ($0\leq z<\infty$), and hence
 we can still employ it as a workhorse. The smaller $|w_2|$,
 the longer term of service is. Note that $w_{de}$ can cross
 the phantom divide $w_{de}=-1$ at
 \be{eq50}
 a_\ast=\exp\left(-\frac{1+w_0}{w_1+w_2}\right)\,,~~~~~~~
 {\rm or~equivalently,}
 ~~~~~~~z_\ast=-1+\exp\left(\frac{1+w_0}{w_1+w_2}\right)\,.
 \ee
 Substituting this $w_{de}$ into the energy
 conservation equation
 $\dot{\rho}_{de}+3H\rho_{de}\left(1+w_{de}\right)=0$, we find that
 \be{eq51}
 \rho_{de}=\rho_{de,0}\,a^{-3(w_1+w_2)/w_2}\left(
 1+w_2\ln a\right)^{3(w_1-w_0 w_2)/w_2^2}\,.
 \ee
 Note that $w_2\not=0$ as mentioned above. Again, in the
 present work, we consider a flat FRW universe containing only
 pressureless matter and dark energy. Substituting Eq.~(\ref{eq51})
 into Friedmann equation, we finally obtain
 \be{eq52}
 E^2=\Omega_{m0}(1+z)^3+\left(1-\Omega_{m0}\right)
 (1+z)^{3(w_1+w_2)/w_2}\left[1-w_2\ln (1+z)
 \right]^{3(w_1-w_0 w_2)/w_2^2}\,.
 \ee
 There are four free parameters in this model,
 namely $\Omega_{m0}$, $w_0$, $w_1$ and $w_2$\,. Note that
 $w_2<0$~is required in Eq.~(\ref{eq49}). By minimizing
 the corresponding total $\chi^2$ in Eq.~(\ref{eq43}), we find
 the best-fit parameters $\Omega_{m0}=0.280$, $w_0=-0.996$,
 $w_1=0.200$, and $w_2=-0.139$, while $\chi^2_{min}=562.254$.
 In Fig.~\ref{fig7}, we present the $68.3\%$ and $95.4\%$
 confidence level contours in the $w_0-w_1$, $w_0-w_2$,
 $w_1-w_2$, $\Omega_{m0}-w_0$, $\Omega_{m0}-w_1$,
 and $\Omega_{m0}-w_2$ planes for the type (II) Pad\'e
 parameterization in Eq.~(\ref{eq47}). Noting Eqs.~(\ref{eq48})
 and (\ref{eq50}), for the best-fit parameters, $w_{de}=-1.438$
 in the very beginning ($a\to 0$, $z\to\infty$), and crossed
 the phantom divide at $a_\ast=0.926$ ($z_\ast=0.076$); today
 ($a=1$, $z=0$), we have $w_{de}=-0.996$; and then $w_{de}$
 will monotonously increase to $+\infty$ at the singularity
 when $a=1328.1$ ($z=-0.999$). This is in fact a so-called
 $w$-singularity~\cite{r35}. According to~\cite{r35}, if this
 $w$-singularity is weak, the spacetime can be extended
 continuously beyond the singularity. From the physical point
 of view, a finite object is not necessarily crushed on
 crossing a weak singularity~\cite{r35}. If our type (II)
 Pad\'e parameterization can cross this $w$-singularity,
 $w_{de}$ will suddenly drop to $-\infty$ when it just crosses
 the singularity, and then it will rapidly increase. Finally,
 $w_{de}\to -1.438$ again in the far future ($a\to\infty$,
 $z\to -1$). We plot this $w_{de}$ as a function of redshift
 $z$ in Fig.~\ref{fig8}, and we can easily see that $w_{de}$
 crossed the phantom divide $w_{de}=-1$. The type (II) Pad\'e
 parameterization works well.

%============================= Fig. 7 =================================

%======================================================================

\vspace{-7mm} % used here just for a more comfortable typesetting

%============================= Fig. 8 =================================

%======================================================================

\vspace{-7mm} % used here just for a more comfortable typesetting

%============================= section 4 ===================================

\section{Conclusion and discussions}\label{sec4}

As is well known, in mathematics, any function could be
 approximated by the Pad\'e approximant. The Pad\'e approximant
 is the best approximation of a function by a rational function
 of given order~\cite{r4}. In fact, the Pad\'e approximant
 often gives better approximation of the function than
 truncating its Taylor series, and it may still work where the
 Taylor series does not converge~\cite{r4}. In the present
 work, we considered the Pad\'e approximant in two issues.
 First, we obtained the analytical approximation of the
 luminosity distance for the flat XCDM model, and found that
 the relative error is fairly small. Second, we proposed
 several parameterizations for the EoS of dark energy based
 on the Pad\'e approximant. They are well motivated from
 the mathematical and physical points of view. We confronted
 these EoS parameterizations with the latest observational
 data, and found that they can work well. In these practices,
 we showed that the Pad\'e approximant could be an useful
 tool in cosmology, and it deserves further investigation.

Here, let us further clarify the physical motivations to use
 the Pad\'e approximant in this work one by one. In the case
 of EoS parameterization, the physical motivation is to avoid
 divergence. As is well known, the singularity is not welcome
 in physics. However, as we mentioned in Sec.~\ref{sec3a}, the
 EoS parameterizations extensively considered in the literature
 diverge in some special cases. For example, the familiar
 parameterization~(\ref{eq20}), $w_{de}=w_0+w_1 z$, diverges
 when redshift $z\to\infty$, and hence this parameterization
 is unsuitable to describe the early universe at high redshift
 $z$. On the other hand, the most popular
 CPL parameterization~(\ref{eq21}), $w_{de}=w_0+w_a (1-a)$,
 diverges when $a\to\infty$ (or $z\to -1$ equivalently). If
 dark energy is phantom-like (its $w_{de}<-1$), as is well
 known, the scale factor $a$ of the universe will diverge
 ($a\to\infty$) in a finite future time (this singularity is
 the well-known big rip). Therefore, the CPL parameterization,
 $w_{de}=w_0+w_a (1-a)$, is unsuitable to describe the
 phantom-like dark energy when the scale factor $a$ is large
 in the future before the big rip. Even if dark energy is
 quintessence-like (its $w_{de}>-1$),  the CPL parameterization
 is also unsuitable to describe the late time universe when
 the scale factor $a$ is very large. On the contrary, as is
 shown in Sec.~\ref{sec3}, the Pad\'e parameterizations
 proposed in this work can easily avoid the divergences
 when $a\to 0$ ($z\to\infty$) and $a\to\infty$ ($z\to -1$),
 unlike the familiar parameterization~(\ref{eq20}) and the most
 popular CPL parameterization~(\ref{eq21}). In particular, as
 mentioned above, the type (I) Pad\'e parameterization proposed
 in Eq.~(\ref{eq31}) is completely free of singularity in the
 whole range $0\leq a<\infty$ ($-1\leq z<\infty$). Therefore,
 using the Pad\'e approximant in EoS parameterization is
 physically motivated, not only mathematically motivated.

Then, let us further clarify the motivation to use the Pad\'e
 approximant in the case of the analytical approximation of the
 luminosity distance. The motivation is to improve the
 calculating efficiency and hence we can save the computing
 time and power. As is well known, the numerical integral
 in the luminosity distance usually employs the algorithms of
 Romberg Integration or Gaussian Quadratures~\cite{r37}. These
 algorithms repeat the calling of the subroutine which
 implements the extended trapezoidal rule and Simpson's
 rule~\cite{r37}, and hence it usually consumes longer time
 than calculating an analytical expression straightforwardly.
 As is shown in e.g.~\cite{r6,r9,r10}, for a given goal
 accuracy, calculating an analytical approximation of the
 luminosity distance is significantly faster than the numerical
 integral employing the algorithms of Romberg Integration or
 Gaussian Quadratures. In particular, as is shown in~\cite{r6},
 the Pad\'e analytical approximation of the luminosity distance
 is the most efficient one in the three existing analytical
 approximations of the luminosity distance for the flat
 $\Lambda$CDM model. This is important. Let us explain in more
 details. When we confront dark energy model with the observational
 data, say, the latest Union2.1 compilation~\cite{r2} which
 consists of 580 SNIa, we need to calculate the theoretical
 luminosity distance for 580 times to compare it with 580 SNIa
 at different redshift $z_{obs}$. When we scan the parameter
 space or run a Monte Carlo simulation, we need to compare dark
 energy model with the observational data for typically $10^6$
 times or even more. So, we have to repeat the calculation of
 the theoretical luminosity distance for $5.8\times 10^8$ times
 or even more. Therefore, even a little improvement in
 calculating efficiency can make a big difference in the
 computing time and power. As mentioned above, an analytical
 approximation of the luminosity distance is significantly
 efficient than the numerical integral employing the algorithms
 of Romberg Integration or Gaussian
 Quadratures~\cite{r6,r9,r10}, while the Pad\'e analytical
 approximation of the luminosity distance is the most efficient
 one~\cite{r6}. Therefore, using the Pad\'e approximant in this
 case is well motivated in fact. Of course, we understand the
 worry about the complicated expressions of $\alpha_i$ and
 $\beta_i$ in Eqs.~(\ref{eq17}) and (\ref{eq18}). However, it
 is not a problem in fact. Noting that the constants $\alpha_i$
 and $\beta_i$ in Eqs.~(\ref{eq17}) and (\ref{eq18}) depend
 only on $w_X$, they are simple numerical values for a given
 $w_X$. For instance, if $w_X=-0.95$, it is easy to find that
 $\alpha_1=1.31874$, $\alpha_2=0.43988$, $\alpha_3=0.0262761$,
 $\beta_1=1.39336$, $\beta_2=0.513621$, $\beta_3=0.0397332$.
 When we write a computer code, these constants are very simple
 for a given $w_X$, and can be written directly or by using
 simply a few lines of code. So, the apparently complicated
 expressions of the constants $\alpha_i$ and $\beta_i$ in
 Eqs.~(\ref{eq17}) and (\ref{eq18}) are not a problem in fact.

\vspace{3.1mm} % used here just for a more comfortable typesetting

%==================== table 1 ====================

 \begin{table}[tbh]
 \begin{center}
 \begin{tabular}{lllll}\hline\hline\\[-3.5mm]
 ~~Model~~~ & CPL & PI & SPI & PII \\[1.2mm] \hline \\[-3.5mm]
 ~~$\chi^2_{min}$ & 562.256 & 562.256 & 562.256 & 562.254 \\
 ~~$k$ & 3 & 4 & 3 & 4 \\
 ~~$\chi^2_{min}/dof$ ~~~ & 0.971081 ~~~ & 0.972761 ~~~
 & 0.971081 ~~~ & 0.972758 ~ \\
 ~~$\Delta$BIC & 0 & 6.36647 & 0 & 6.36447 \\
 ~~$\Delta$AIC & 0 & 2 & 0 & 1.998 \\
 ~~Rank & 1 & 3 & 1 & 2 \\[1.2mm] \hline\hline
 \end{tabular}
 \end{center}
 \caption{\label{tab1} Comparing CPL parameterization with
 type (I) Pad\'e parameterization (PI), simplified type
 (I) Pad\'e parameterization (SPI), and type (II) Pad\'e
 parameterization (PII). Note that CPL parameterization
 has been chosen to be the fiducial model when we calculate
 $\Delta$BIC and $\Delta$AIC. See the text for details.}
 \end{table}

%=================================================

\vspace{3.1mm} % used here just for a more comfortable typesetting

It is of interest to compare our Pad\'e EoS parameterizations
 with the well-known CPL parameterization. Since these models
 have different free parameters and the correlations between
 model parameters are fairly different, it is not suitable to
 directly compare their confidence level contours. Instead, as
 in the literature, it is more appropriate to compare them from
 the viewpoint of goodness-of-fit. A conventional criterion for
 model comparison in the literature is $\chi^2_{min}/dof$, in
 which the degree of freedom $dof={\cal N}-k$, while $\cal N$
 and $k$ are the number of data points and the number of free
 model parameters, respectively. On the other hand, there
 are other criterions for model comparison in the literature.
 The most sophisticated criterion is the Bayesian evidence
 (see e.g.~\cite{r38} and references therein). However,
 the computation of Bayesian evidence usually consumes a
 large amount of time and power. As an alternative, one can
 consider some approximations of Bayesian evidence, such as
 the so-called Bayesian Information Criterion (BIC) and
 Akaike Information Criterion (AIC). The BIC is defined
 by~\cite{r39}
 \be{eq53}
 {\rm BIC}=-2\ln{\cal L}_{max}+k\ln {\cal N}\,,
 \ee
 where ${\cal L}_{max}$ is the maximum likelihood. In the
 Gaussian cases, $\chi^2_{min}=-2\ln{\cal L}_{max}$. So, the
 difference in BIC between two models is given by
 $\Delta{\rm BIC}=\Delta\chi^2_{min}+\Delta k \ln {\cal N}$.
 The AIC is defined by~\cite{r40}
 \be{eq54}
 {\rm AIC}=-2\ln{\cal L}_{max}+2k\,.
 \ee
 The difference in AIC between two models is
 given by $\Delta{\rm AIC}=\Delta\chi^2_{min}+2\Delta k$.
 As is well known, the corresponding $E\equiv H/H_0$ for CPL
 parameterization~(\ref{eq21}) is given by
 (see e.g.~\cite{r12})
 \be{eq55}
 E(z)=\left[\Omega_{m0}(1+z)^3
 +\left(1-\Omega_{m0}\right)(1+z)^{3(1+w_0+w_a)}
 \exp\left(-\frac{3 w_a z}{1+z}\right)\right]^{1/2}\,.
 \ee
 There are three independent parameters. By minimizing the
 corresponding total $\chi^2$ in Eq.~(\ref{eq43}), we find the
 best-fit parameters $\Omega_{m0}=0.280$, $w_0=-0.995$ and
 $w_a=-0.074$, while $\chi^2_{min}=562.256$. In Table~\ref{tab1}, we
 present $\chi^2_{min}/dof$, $\Delta$BIC and $\Delta$AIC for
 CPL parameterization, type (I) Pad\'e parameterization (PI),
 simplified type (I) Pad\'e parameterization (SPI), and type
 (II) Pad\'e parameterization (PII). Note that CPL parameterization
 has been chosen to be the fiducial model when we calculate
 $\Delta$BIC and $\Delta$AIC. From Table~\ref{tab1}, we see
 that the rank of models is coincident in all the three
 criterions, namely, $\chi^2_{min}/dof$, BIC and AIC. The CPL
 parameterization is slightly better than type (I) and (II)
 Pad\'e parameterizations given in Eqs.~(\ref{eq31}) and
 (\ref{eq47}). However, the simplified type (I) Pad\'e
 parameterization given in Eq.~(\ref{eq46}) is as good as CPL
 parameterization.

Some remarks are in order. First, it is important to obtain an
 analytical approximation of the luminosity distance for dark
 energy models with variable EoS $w_{de}$. We have tried the
 dark energy model with CPL EoS $w_{de}=w_0+w_a(1-a)$ but
 failed because the relative error is unacceptably large.
 This issue deserves further attempts, and we leave it as an
 open question. Second, it is also of interest to find an
 analytical approximation of the luminosity distance for a
 non-flat FRW universe. Even, one can further consider an
 inhomogeneous or anisotropic universe, say, a
 Lemaitre-Tolman-Bondi (LTB) universe. Third, since the angular
 diameter distance $d_A=d_L(1+z)^{-2}$, the corresponding
 analytical approximation of the angular diameter distance is
 also ready. Fourth, in both issues of the luminosity distance
 and EoS parameterization, one can of course use a Pad\'e
 approximant of higher order to improve the accuracy, rather
 than just order $(3,\,3)$ in $d_L$, or order $(1,\,1)$ in
 EoS parameterization as we done. However, this will bring the
 drawback of heavier computation, or too many free model
 parameters. Fifth, in Sec.~\ref{sec3b}, motivated
 by Eq.~(\ref{eq33}) and the best-fit parameters, we can
 consider other two simplified versions of the type (I)
 Pad\'e parameterization, namely
 \be{eq56}
 w_{de}=\frac{-1+w_a(1-a)}{1+w_b(1-a)}\,,
 \ee
 by setting $w_0=-1$, or
 \be{eq57}
 w_{de}=\frac{w_0-w_b(1-a)}{1+w_b(1-a)}\,,
 \ee
 by setting $w_a=-w_b$, while $-1<w_b<0$ still holds. They have
 only two free parameters, and different interesting behaviors.
 Sixth, although there exists a so-called
 $w$-singularity~\cite{r35} in the type (II)
 Pad\'e parameterization, it might be not so serious. According
 to~\cite{r35}, if this $w$-singularity is weak, the spacetime
 can be extended continuously beyond the singularity. From
 the physical point of view, a finite object is not necessarily
 crushed on crossing a weak singularity~\cite{r35}. So, it is
 of interest to study whether this type of $w$-singularity is
 weak, and we leave it to the future works. Seventh, in fact,
 the present work is not the first one using the Pad\'e
 approximant in cosmology. We refer to e.g.~\cite{r36,r41} (see
 also~\cite{r6,r7}) for the previous relevant works. In these
 works, the Pad\'e approximant has been used in the slow-roll
 inflation, the reconstruction of the scalar field potential
 from SNIa, the data fitting of luminosity distance, a special
 EoS parameterization with respect to redshift $z$, and the
 cosmological perturbation in LSS. Anyway, the issues discussed
 in the present work are different from the previous works in
 the literature. Eighth, since our type (I) Pad\'e parameterization
 in Eq.~(\ref{eq31}) is well motivated from mathematics, and
 completely free of singularity in the whole range $0\leq a<\infty$
 ($-1\leq z<\infty$) unlike the CPL parameterization, we
 recommend the community to use it or its variants in the
 relevant works. Finally, as shown in this work, the Pad\'e
 approximant is useful. It is of interest to consider its
 other applications in cosmology.

%============================= acknowledgements ===================================

\section*{ACKNOWLEDGEMENTS}
We thank the anonymous referee for quite useful comments and
 suggestions, which helped us to improve this work. We are
 grateful to Professors Rong-Gen~Cai, Shuang~Nan~Zhang and
 Tong-Jie~Zhang for helpful discussions. We also thank
 Minzi~Feng, as well as Long-Fei~Wang, Xiao-Jiao~Guo, Zu-Cheng~Chen
 and Jing~Liu, for kind help and discussions. This work was
 supported in part by NSFC under Grants No.~11175016 and
 No.~10905005, as well as NCET under Grant No.~NCET-11-0790,
 and the Fundamental Research Fund of Beijing Institute of
 Technology.

\renewcommand{\baselinestretch}{1.12}

%============================= references ==================================

\end{document}